\documentclass[twocolumn]{IEEEtran}
\IEEEoverridecommandlockouts
\usepackage{cite}
\usepackage{amsmath,amssymb,amsfonts}
\usepackage{algorithm,algorithmic}
\usepackage{graphicx,subfigure}
\usepackage{textcomp}
\usepackage{multirow}
\usepackage{float,color}
\usepackage{bm}
\def\BibTeX{{\rm B\kern-.05em{\sc i\kern-.025em b}\kern-.08em
    T\kern-.1667em\lower.7ex\hbox{E}\kern-.125emX}}

\newcommand{\st}{\text{s.t.}}

\newcommand{\snr}{\text{SNR}}
\newcommand{\Jcal}{\mathcal{J}}

\renewcommand{\baselinestretch}{0.99}

\begin{document}

\title{Parameter-free $\ell_p$-Box Decoding of LDPC Codes\\
\thanks{Qiong Wu, Fan Zhang and Hao Wang are with School of Information Science and Technology, ShanghaiTech University, Shanghai, P.R. China (email: \{wuqiong, zhangfan4, wanghao1\}@shanghaiTech.edu.cn).}
\thanks{Jun Lin is with School of Electronic Science and Engineering, Nanjing University, P.R. China (email: jlin@nju.edu.cn).}
\thanks{Yang Liu is with Department of Electrical and Computer Engineering, Lehigh University, PA, USA (email: liuocean613@gmail.com).}
}

\author{
\IEEEauthorblockN{Qiong Wu, Fan Zhang, Hao Wang, Jun Lin and Yang Liu\\}
}
    
%
%

\maketitle

\begin{abstract}
The Alternating Direction Method of Multipliers (ADMM) decoding of Low Density Parity Check (LDPC) codes has received many attentions due to its excellent performance at the error floor region. In this paper, we develop a parameter-free decoder based on Linear Program (LP) decoding  by replacing the binary constraint with the intersection of a box and an $\ell_p$ sphere. An efficient $\ell_2$-box ADMM is designed to handle this model in a distributed fashion. Numerical experiments demonstrates that our decoder attains better adaptability to different Signal-to-Noise Ratio and channels.
\end{abstract}


\section{Introduction}

Low Density Parity Check (LDPC) codes have been widely applied nowadays\cite{LDPC}.
Although various decoding methods have been developed for LDPC, e.g. Belief Propagation (BP) and Min-Sum (MS) \cite{EFAF, LDPC}, nearly all of the existing solutions are approximation-based (not truly Maximum Likelihood solution) and consequently suffer an early error floor, especially in the case of high Signal-to-Noise Ratio (SNR). Therefore, it is important yet challenging to develop an error-floor-free decoding scheme for LDPC with more accurate decoding performance.

For binary LDPC codes  over symmetric channels, Feldman introduced a relaxed version \cite{DECC, ULPT} of Maximum Likelihood (ML) decoding problem which can be interpreted as a Linear Program (LP) decoding, whose certificate of correctness (ML certificate) and  performance  are reported in \cite{OLCL}.  Compared with other decoding methods, such as BP and MS, LP decoding can reduce the error floor\cite{MPAA}. Many algorithms have been proposed for LP decoding. The most related work is the Alternating Direction Method of Multipliers (ADMM)  \cite{ADMM}  introduced in  \cite{DMFL}, which can be easily implemented in a distributed way.  

In order to obtain the ideal decoding performance, Liu et al. \cite{SPBP} developed a penalized-LP decoding model by adding different types of penalty terms in the objective to drive the solution away from 0.5, and reported that the $\ell_2$ penalty term generally achieved better error performance.  
However, one remaining hurdle is the difficulty in properly choosing the penalty parameter. The effect of the penalty parameter is two-fold.  First,  the penalty parameter closely affects the performance of the algorithm and an improper value can make the algorithm extremely inefficient. Most importantly,  the optimal solution of the penalized-LP decoding problem may vary for different penalty parameter values. As a result, the decoder often finds  solutions with low accuracy.  It is worth noting that the penalized-LP formulation in fact approximates the original problem via penalizing the violation of the binary constraint.  Hence its optimal solution is rarely  the optimal solution of the original LP decoding problem. 

In this paper,  we apply a parameter-free continuous optimization model \cite{AVFF}, which is an \underline{\textbf{\emph{exact}}} reformulation of the binary LP decoding.  The binary constraint in the LP decoding is replaced by the intersection of a box and an $\ell_p$ sphere.  We also design an efficient ADMM solver, which can be parallelized to reduce the computational cost.  Simulation results demonstrate that the proposed model along with the ADMM solver can further  bring down  the error floor for large SNRs. 

\section{Background}\label{sec.back}

We consider a binary linear LDPC code $C$ of length $N$, which can be defined by an $M \times N$ parity check matrix $\bm{H}$. Each column of the parity check matrix corresponds to a codeword symbol, indexed by $\mathcal{I}:=\{1,2,...,N\} $. Each row of the parity check matrix corresponds to a check, which specifies a subset of codeword symbols that add to 0 (modulo 2), indexed by $\mathcal{J}:=\{1,2,...,M\} $. The neighborhood of check $j$, denoted as $\mathcal{N} (j)$, is the set of variables that check $j$ constrains to add to 0. That is \cite{GECC}, $\mathcal{N} (j) = \{ i \in \mathcal{I}:\bm{H}_{j,i} = 1 \}$.

For a binary linear code transmitted over a symmetric memoryless channel, let $\mathcal{X} = \{0, 1\}$ be the input space, $\mathcal{Y}$ be the output space and $P(\bm{y}|\bm{x})$ denotes the probability that $\bm{y} \in \mathcal{Y}$ is received if the codeword $\bm{x} \in \mathcal{X}$ is sent over the channel.  Let $\mathcal{C}$ be the set of possible codewords. By Bayes' rule,  the decoding problem can be modeled   as $\arg \max_{\bm{x} \in \mathcal{C}} P(\bm{y}|\bm{x})$. In particular, the Maximum Log-likelihood (ML) decoding problem \cite{DECC,ULPT} 
takes the form 
$
\hat{\bm{x}}= \arg \max_{\bm{x} \in \mathcal{C}} P(\bm{y}|\bm{x}) = \arg \min_{\bm{x} \in \mathcal{C}} \bm{\gamma}^T \bm{x},
$
where $\bm{\gamma}$ is a length-$N$ vector of Log-likelihood Ratios (LLRs) with $\bm{\gamma}_i = \log (P(y_i|0)/P(y_i|1) )$.

Some work \cite{DMFL,ANLP} has been done in the past decades focusing on the properties of the convex hull of all possible codewords. ML decoding can be formulated as the  problem of minimizing $\bm{\gamma}^T \bm{x}$ over the convex hull---an Integer Programming (IP) decoding problem that is NP-hard in general\cite{OTII}.
Denote the subset of coordinates of $\bm{x}$  participating in the $j$th check as the matrix $\bm{P}_j$, so $\bm{P}_j$ is a binary $d \times N$ matrix consisting of $d$ components participating in the $j$th check.  From \cite{DMFL},   each local codeword constraint can be relaxed to satisfy  $\bm{P}_j \bm{x} \in \mathbb{PP}_d$, where $\mathbb{PP}_d$ is the specific expression of the convex hull of all possible codewords. 
The LP decoding problem is then derived in \cite{DMFL}, by relaxing the binary constraints
\begin{equation}\label{LP}
 	 \min_{\bm{x}}\quad \bm{\gamma}^T \bm{x} \quad
 	 \st \quad  \bm{x} \in [0,1]^N,\quad \bm{P}_j \bm{x} \in \mathbb{PP}_d , \forall j \in \mathcal{J},
\end{equation}
where  $[0, 1]^N$ is the $N$-dimensional box. The binary constraint $\bm{x} \in \{0,1\}^N$ in ML decoding is relaxed to $\bm{x} \in [0,1]^N$ now. 


The simulation results in \cite{DMFL} suggest that an ideal decoder should perform like BP at low SNRs. To achieve this, a penalized-LP decoding is proposed in  \cite{SPBP}  by  penalizing the fractional solutions, that is, the objective of \eqref{LP} is replaced with $\bm{\gamma}^T \bm{x} + \sum_{i=1}^N g(x_i)$,
where the penalty function $g : [0, 1]\rightarrow \mathbb{R} \cup \{\pm \infty \} $ has three options, namely,  $\ell_1$, $\ell_2$ and $\log$ penalty. That is, 
$g_1(x) = −\alpha|x - 0.5|$,
$g_2(x) = −\alpha(x - 0.5)^2$,
$g_3(x) = −\alpha\log(|x - 0.5|)$, with penalty parameter $\alpha>0$.  
It is reported that the $\ell_2$ penalty has the best error performance and the fastest convergence among the options. Other improved penalty functions are proposed and experimented in \cite{IPFO}.

\section{ $\ell_p$-box decoding}

In this section, we propose a new formulation for LP decoding. Compared to the existing formulations suitable to decentralized processing, where mixed integer formulation are usually involved, our proposed formulation is to design a continuous and exact reformulation of the LP decoding to circumvent using the traditional method to solve the IP problem, such as branch and bound  and Lagrangian relaxation. 
The major technique used in our model is the $\ell_p$-box recently proposed in \cite{AVFF}   for solving IP problems.
The main idea of this technique is to replace the binary constraint   by the intersection of a box and an $\ell_p$ sphere with $p \in (0, \infty)$:
\begin{equation}\label{lpbox}
\bm{x} \in \{0,1\}^N  \Leftrightarrow \bm{x} \in [0,1]^N \cap \{ \bm{x}: \|\bm{x} - \tfrac{1}{2} \mathbf{1}_N\|_p^p = \tfrac{N}{2^p} \},
\end{equation}
where $\mathbf{1}_N$ is the $N$-dimensional vector filled with all $1$s. 

Now consider this technique in the LP decoding problem \eqref{LP}.      Besides the existing box constraint $[0,1]^N$,  we only need an additional $\ell_p$-sphere constraint to enforce the solution to be binary, yielding the $\ell_p$-box   decoding problem:
\begin{equation}\label{lpbox_ADMM}
\begin{aligned}
 	 \min_{\bm{x}}\ &\bm{\gamma}^T \bm{x}\\
 	 \st\  & \bm{x} \in [0,1]^N, \|\bm{x} -\tfrac{1}{2}\mathbf{1}_N\|_p^p = \tfrac{N}{2^p}, \bm{P}_j \bm{x} \in \mathbb{PP}_d , \forall j \in \mathcal{J}.
\end{aligned}
\end{equation}

For the penalized-LP decoding, the choice of parameters is critical to the solution performance and, unfortunately, an appropriate choice can be very intricate to obtain. If the penalty parameter is set too large, the algorithm may quickly converge to a binary point far from the global solution. On the other hand, 
for a too small penalty parameter, the algorithm may converge  to a fractional point. 
Compared to the parameterized penalized-LP formulation, there is no parameter involved in \eqref{lpbox_ADMM}, rescuing users from adjusting the penalty parameters according to SNR. 
It should be noticed that  \eqref{lpbox_ADMM} is equivalent to 
the binary decoding problem. Therefore, the optimal solution of the binary decoding problem is 
the global optimal solution of \eqref{lpbox_ADMM}. Furthermore, \eqref{lpbox_ADMM} is a continuous problem, and thus can be attacked by various continuous optimization algorithms.

There are many options for the selection of $p$, and \eqref{lpbox_ADMM} becomes nonsmooth with $p\in(0,1]$. 
In this paper, we use $p=2$ 
to keep the projection of a point onto the sphere simple.

\section{$\ell_2$-box ADMM}\label{sec.admm}
In this section, we design an efficient algorithm for solving the proposed $\ell_2$-box decoding problem by  incorporating the $\ell_2$-box  technique in the ADMM framework.

As discussed in previous section,  we set $p=2$.
%
In order to apply the ADMM, we introduce two sets of auxiliary variables $\bm{y}$ and $\bm{z}$ to decouple the box and the $\ell_2$ sphere, resulting in an $\ell_2$-box decoding problem:
\begin{equation}\label{l2box_ADMM}
\begin{aligned}
 	 \min_{\bm{x, y, z}}\quad &\bm{\gamma}^T \bm{x}\\
 	 \st\quad 
		&\bm{y}= \bm{x},\quad \|\bm{y}-\tfrac{1}{2}\mathbf{1}_N\|_2^2 = \tfrac{N}{4},\quad \bm{x} \in [0,1]^N\\
		&\bm{z}_j= \bm{P}_j \bm{x} ,\quad \bm{z}_j\in \mathbb{PP}_d ,\quad \forall j \in \mathcal{J},
\end{aligned}
\end{equation}
where $\bm{y}$ is a $N$-dimensional vector and $\bm{z}_j \in \mathbb{PP}_d $ for all $j \in \mathcal{J}$.
The ADMM   works with the augmented Lagrangian
$
L_{\mu_1,\mu_2}(\bm{x},\bm{y},\bm{z},\bm{\lambda}_1,\bm{\lambda}_2) = \bm{\gamma}^T \bm{x} + \sum\limits_{j\in \mathcal{J}} \bm{\lambda}_{1,j}^T(\bm{P}_j \bm{x} -\bm{z}_j) 
+\frac{\mu_1}{2}\sum\limits_{j\in \mathcal{J}}\|\bm{P}_j \bm{x} -\bm{z}_j\|_2^2 + \bm{\lambda}_2^T(\bm{x} -\bm{y})  +\frac{\mu_2}{2}\|\bm{x} -\bm{y}\|_2^2
$
with constants $\mu_1$ and $\mu_2$ all positive.
Here $\bm{\lambda}_1$ is the dual variable associated with the constraint $\bm{z}_j= \bm{P}_j x$ and its dimension is the cardinality of $|\mathcal{J}|$, 
and the $N$-dimensional $\bm{\lambda}_2$ is the dual variable associated with the coupling constraint $\bm{y}= \bm{x}$.


Let 
$\mathcal{X}$, $\mathcal{Y}$  and $\mathcal{Z}$ denote the feasible regions for variables $\bm{x}$, $\bm{y}$ and $\bm{z}$ respectively, i.e., $\mathcal{X} =\mathcal{Y}= [0,1]^N$ and
\begin{equation*}
  \begin{array}{rll}
\mathcal{Z} = & \underbrace{\mathbb{PP}_{|\mathcal{N}(j)|} \times ...\times \mathbb{PP}_{|\mathcal{N}(j)|}}_{number\ of \ checks\  \mathcal{J}}
  \end{array} .
\end {equation*}
The ADMM iteration then can be elaborated as follows 
{\small
\begin{equation*}
\bm{x}\text{-update}: \ \bm{x}^{k+1} = \arg\min_{\bm{x} \in \mathcal{X}} L_{\mu_1,\mu_2}(\bm{x},\bm{y}^k,\bm{z}^k,\bm{\lambda}_1^k,\bm{\lambda}_2^k),
\end{equation*}
\begin{equation*}
(\bm{y},\bm{x})\text{-update}:  \left\{\begin{split}
&\bm{y}^{k+1} = \arg\min_{\bm{y} \in \mathcal{Y}} L_{\mu_1,\mu_2}(\bm{x}^{k+1},\bm{y},\bm{z}^k,\bm{\lambda}_1^k,\bm{\lambda}_2^k) \\ \ & \st \ \|\bm{y}-\tfrac{1}{2}\mathbf{1}_N\|_2^2 = \tfrac{N}{4} \\
&\bm{z}^{k+1} = \arg\min_{z \in \mathcal{Z}} L_{\mu_1,\mu_2}(\bm{x}^{k+1},\bm{y}^k,\bm{z},\bm{\lambda}_1^k,\bm{\lambda}_2^k)\\  \  & \st \ \bm{z}_j\in \mathbb{PP}_d ,\quad \forall j \in \mathcal{J},
\end{split}\right.
\end{equation*}
\begin{equation*}
(\bm{\lambda}_1,\bm{\lambda}_2)\text{-update}:  \left\{\begin{split}
&\bm{\lambda}_{1,j}^{k+1} = \bm{\lambda}_{1,j}^{k} + \mu_1(\bm{P}_j\bm{x}^{k+1} - \bm{z}_j^{k+1}) ,\  \forall j \in \mathcal{J}\\
&\bm{\lambda}_2^{k+1} = \bm{\lambda}_2^{k} + \mu_2(\bm{x}^{k+1} - \bm{y}^{k+1}).
\end{split}\right.
\end{equation*}
}
Next, we discuss the solution of the ADMM subproblems.

(a) $\bm{x}$-update. The $\bm{x}$-update is to minimize $L_{\mu_1,\mu_2}(\bm{x},\bm{y},\bm{z},\bm{\lambda}_1,\bm{\lambda}_2)$ subject to $\bm{x}\in[0,1]^N$ with fixed $(\bm{y},\bm{z},\bm{\lambda}_1,\bm{\lambda}_2)$,
The $i$th component of the solution of the $\bm{x}$-update subproblem can be given explicitly by 
$
\bm{x}_i^{k+1} = \Pi_{[0,1]^N}( \tfrac{   d^k- \gamma_i - \bm{\lambda}_{2,i}^k +\mu_2  \bm{y}_i^k     }{\mu_1|\mathcal{N}(i)|+\mu_2}   ).
$
Here $d^k = \sum_{j \in \mathcal{N}(i)}   \mu_1 (\bm{P_j}^T\bm{z}_j^k)^{(i)} - (\bm{P_j}^T\bm{\lambda}_{1,j}^k)^{(i)} $
where $(\bm{P_j}^T\bm{z}_j^k)^{(i)}$ denotes the $i$th component of $\bm{P_j}^T\bm{z}_j^k$, and so forth for $(\bm{P_j}^T\bm{\lambda}_{1,j}^k)^{(i)}$, $\bm{x}_i$, $\bm{\lambda}_{2,i}$ and $\bm{y}_i$.

Define  $\bm{P} = \sum_{j\in \mathcal{J}} \bm{P}_j^T\bm{P}_j$ and $ \Pi_{[0,1]^N} $ as the projection operator  onto the $N$-dimensional box $[0, 1]^N$.
Note that for any $j$, $\bm{P} = \sum_{j\in \mathcal{J}} \bm{P}_j^T\bm{P}_j$ is an $N\times N$ diagonal binary
matrix with non-zero entries at $(i, i)$ if and only if $i$ participates in the $j$th parity check, i.e., $i \in \mathcal{N}(j)$.
This implies that $\bm{P} = \sum_{j\in \mathcal{J}} \bm{P}_j^T\bm{P}_j$ is a diagonal matrix with the $(i, i)$th entry equal to $|\mathcal{N}(i)|$. Hence $\bm{P}^{-1}$ is diagonal with $1/|\mathcal{N}(i)|$ as the $i$th diagonal entry. See details of this calculation in \cite{DMFL}.

(b) $(\bm{y},\bm{z})$-update. Notice that  the two variables $\bm{y}$ and $\bm{z}$ are independent to each other, thus they  can be  updated separately. 
By fixing $(\bm{x},\bm{z},\bm{\lambda}_1,\bm{\lambda}_2)$,  the $\bm{y}$-update subproblem is given by 
\begin{equation}\label{prob.y}
\begin{aligned}
&\bm{y}^{k+1} = \arg\min_{\bm{y} \in \mathcal{Y}} L_{\mu_1,\mu_2}(\bm{x}^{k+1},\bm{y},\bm{z}^k,\bm{\lambda}_1^k,\bm{\lambda}_2^k),\\ \ &\st \ \|\bm{y}-\tfrac{1}{2}\mathbf{1}_N\|_2^2 = \tfrac{N}{4}.
\end{aligned}
\end{equation}
Thus, by eliminating the constants, \eqref{prob.y} can be written as 
\begin{equation*}
\min_{\bm{y}}\   - \left( \mu_2(\bm{x}^{k+1}-\tfrac{1}{2}\mathbf{1}_N) + \bm{\lambda}_2^{k}  \right)^T \bm{y} \quad \st \ \|\bm{y}-\tfrac{1}{2}\mathbf{1}_N\|_2^2 = \tfrac{N}{4},
\end{equation*}
 which has an explicit solution:
$
\bm{y}^{k+1} = \tfrac{\mu_2(\bm{x}^{k+1}-\tfrac{1}{2}\mathbf{1}_N) + \bm{\lambda}_2^{k}}{\|\mu_2(\bm{x}^{k+1}-\tfrac{1}{2}\mathbf{1}_N) + \bm{\lambda}_2^{k}\|} \times \tfrac{\sqrt{N}}{2} + \tfrac{1}{2}.
$

For updating $\bm{z}$,  the subproblem is to solve 
\begin{equation}
\min_{\bm{z}_j} \    \tfrac{\mu_1}{2}\|\bm{P}_j \bm{x}^{k+1} -\bm{z}_j\|_2^2 - \bm{\lambda}^k_{1,j} \bm{z}_j, \  \st \ \bm{z}_j \in \mathbb{PP}_d,
\end{equation}
for each $j \in \mathcal{J}$.   The solution of this subproblem can be given explicitly  \cite{SPBP}:
$
\bm{z}^{k+1}_j =\Pi_{\mathbb{PP}_d} (\bm{P}_j\bm{x}^{k+1} + \tfrac{1}{\mu_1} \bm{\lambda}^k_{1,j} ), \ \forall j \in\mathcal{J},
$
where $ \Pi_{\mathbb{PP}_d} $ is the projection operator onto the codeword polytope.

(c) $(\bm{\lambda}_1,\bm{\lambda}_2)$-update. Note that $\bm{\lambda}_1$ and $\bm{\lambda}_2$ are also independent to each other, and thus can be updated separately:
\begin{equation}
\begin{aligned}
&\bm{\lambda}_{1,j}^{k+1} = \bm{\lambda}_{1,j}^{k} + \mu_1(\bm{P}_j\bm{x}^{k+1} - \bm{z}_j^{k+1}), \  \forall j\in\mathcal{J}\\
&\bm{\lambda}_2^{k+1} = \bm{\lambda}_2^{k} + \mu_2(\bm{x}^{k+1} - \bm{y}^{k+1}).
\end{aligned}
\end{equation}

The entire $\ell_2$-box ADMM decoding algorithm is stated in Algorithm~\ref{alg.admm}. 

\begin{algorithm}[htbp]
	\caption{$\ell_2$-box ADMM decoding}
	\label{alg.admm}
	\begin{algorithmic}[1]
		\STATE Given a $M \times N$ parity check matrix $\bm{H}$, and parameters $\mu_1>0$, $\mu_2>0$, tolerance $\epsilon>0$.
		\STATE Construct the log-likelihood vector $\bm{\gamma}$ and the $d \times N$
		matrix $\bm{P}_j $ for all $j \in \Jcal$.
		\STATE Initialize $\bm{y}$, $\bm{\lambda}_2$, $\bm{z}_j$ and $\bm{\lambda}_{1,j}$ for all $j \in \Jcal$.
		\STATE Set $k$ = 0.  
		\REPEAT
		\STATE \textbf{Update $\bm{x}$}:\ $\bm{x}_i^{k+1} = \Pi_{[0,1]^N}( \frac{   d^k- \gamma_i - \bm{\lambda}_{2,i}^k +\mu_2  \bm{y}_i^k     }{\mu_1|\mathcal{N}(i)|+\mu_2}  ).
			$
		
		\STATE \textbf{Update $\bm{y}$}:\ $ \bm{y}^{k+1} = \frac{\mu_2(\bm{x}^k -\tfrac{1}{2}\mathbf{1}_N) + \bm{\lambda}_2^k}{\|\mu_2(\bm{x}^k -\tfrac{1}{2}\mathbf{1}_N) + \bm{\lambda}_2^k\|}\times \frac{\sqrt{N}}{2} + \tfrac{1}{2}$
		
		\STATE \textbf{Update $\bm{z}$}:\\ \ for all $j\in \mathcal{J}$ do
		$\bm{z}^{k+1}_j =\Pi_{\mathbb{PP}_d} (\bm{P}_j\bm{x}^{k+1} + \frac{1}{\mu_1} \bm{\lambda}^k_{1,j} )$

		\STATE \textbf{Update $\bm{\lambda}_1$}:\\ \  for all $j\in \mathcal{J}$ do $ \bm{\lambda}_{1,j}^{k+1} = \bm{\lambda}_{1,j}^{k} + \mu_1(\bm{P}_j\bm{x}^{k+1} - \bm{z}_j^{k+1})$
		
		\STATE \textbf{Update $\bm{\lambda}_2$}:\\ \ $ \bm{\lambda}_2^{k+1} = \bm{\lambda}_2^{k} + \mu_2(\bm{x}^{k+1} - \bm{y}^{k+1}) $
		
		\STATE  Set  $k = k+1$
		
		\UNTIL{{\small$\max\limits_j \|\bm{P}_j\bm{x}^{k+1} - \bm{z}_j^{k+1} \|_{\infty} < \epsilon $ and $\|\bm{x}^{k+1} - \bm{y}^{k+1}\|_{\infty}<\epsilon$}}
		\STATE \textbf{return} $\bm{x}$
	\end{algorithmic}
\end{algorithm}

\section{Numerical Experiments}\label{sec.test}

In this section, we design numerical experiments to test the proposed decoding model and algorithm. 
We use the experimental framework of Liu\cite{SPBP} in our test, 
and design a simulator that the $[2640,1320]$ ``Margulis" code \cite{CCCM}, which is known for its strong error-floor,  with randomly added noise transmitting over an Additive White Gaussian Noise (AWGN) channel.

Many acceleration techniques exist for ADMM \cite{FADO}, and the $\ell_2$-penalized ADMM LP decoder in our comparison is in fact an accelerated version \cite{FADO}. Our algorithm, however, is simply the basic ADMM without using acceleration, since our primary focus is on the effectiveness of our proposed decoding model. The ADMM parameters in our implementation are roughly tuned to ensure the algorithm achieve an acceptable performance.

%

\subsection{Effectiveness of the parameter-free model}
To emphasize the advantage of our new formulation, we first show  for the $\ell_2$-penalized ADMM LP decoder how the penalty parameter value could affect the decoding accuracy and running time.

\begin{figure}[htbp]
\centering 
\subfigure[WER versus $\alpha$]{ \label{fig:para:alpha_WER} 
\includegraphics[width=1.65in]{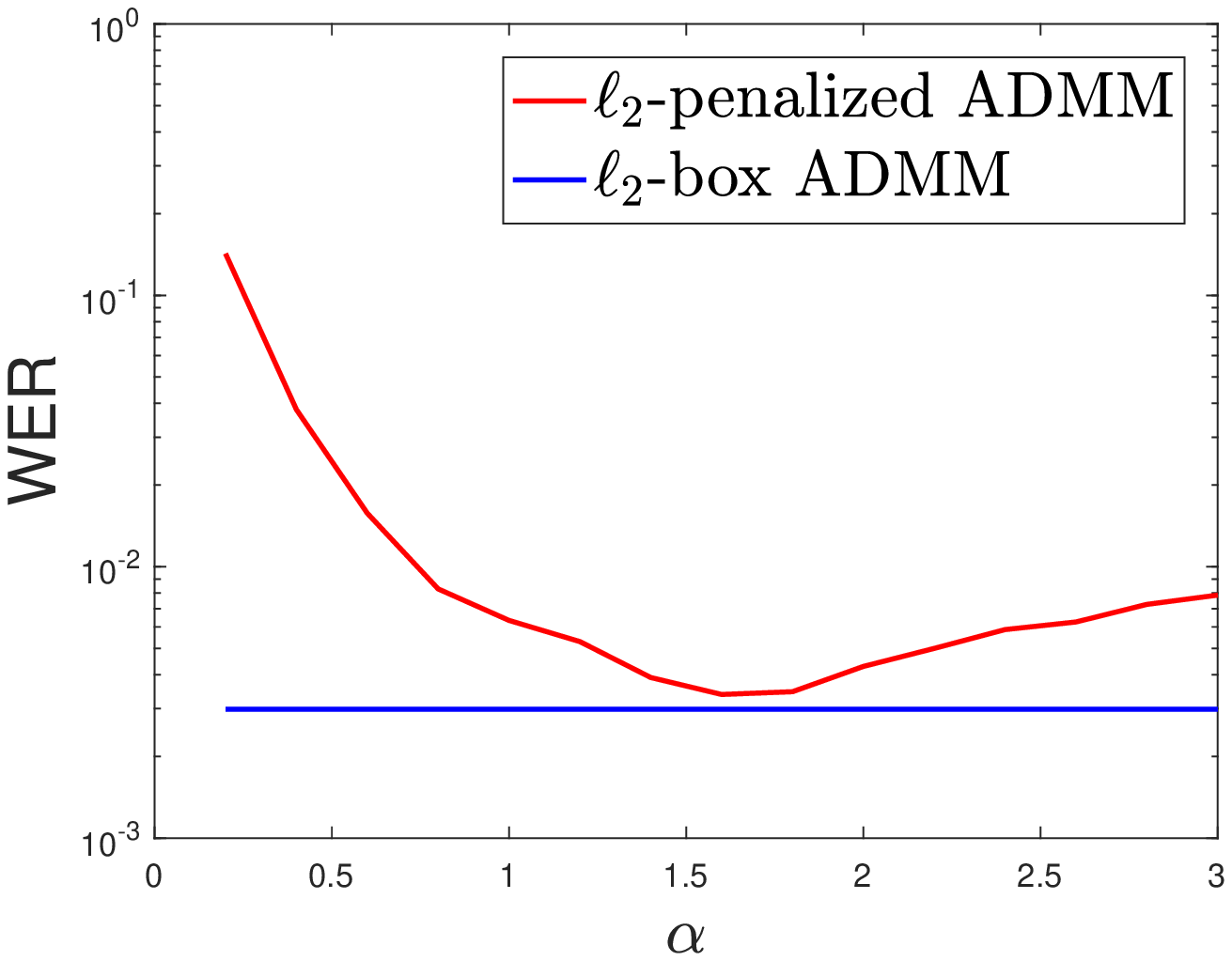}} 
\hspace{0in} 
\subfigure[Running time versus $\alpha$]{ \label{fig:para:alpha_time} 
\includegraphics[width=1.65in]{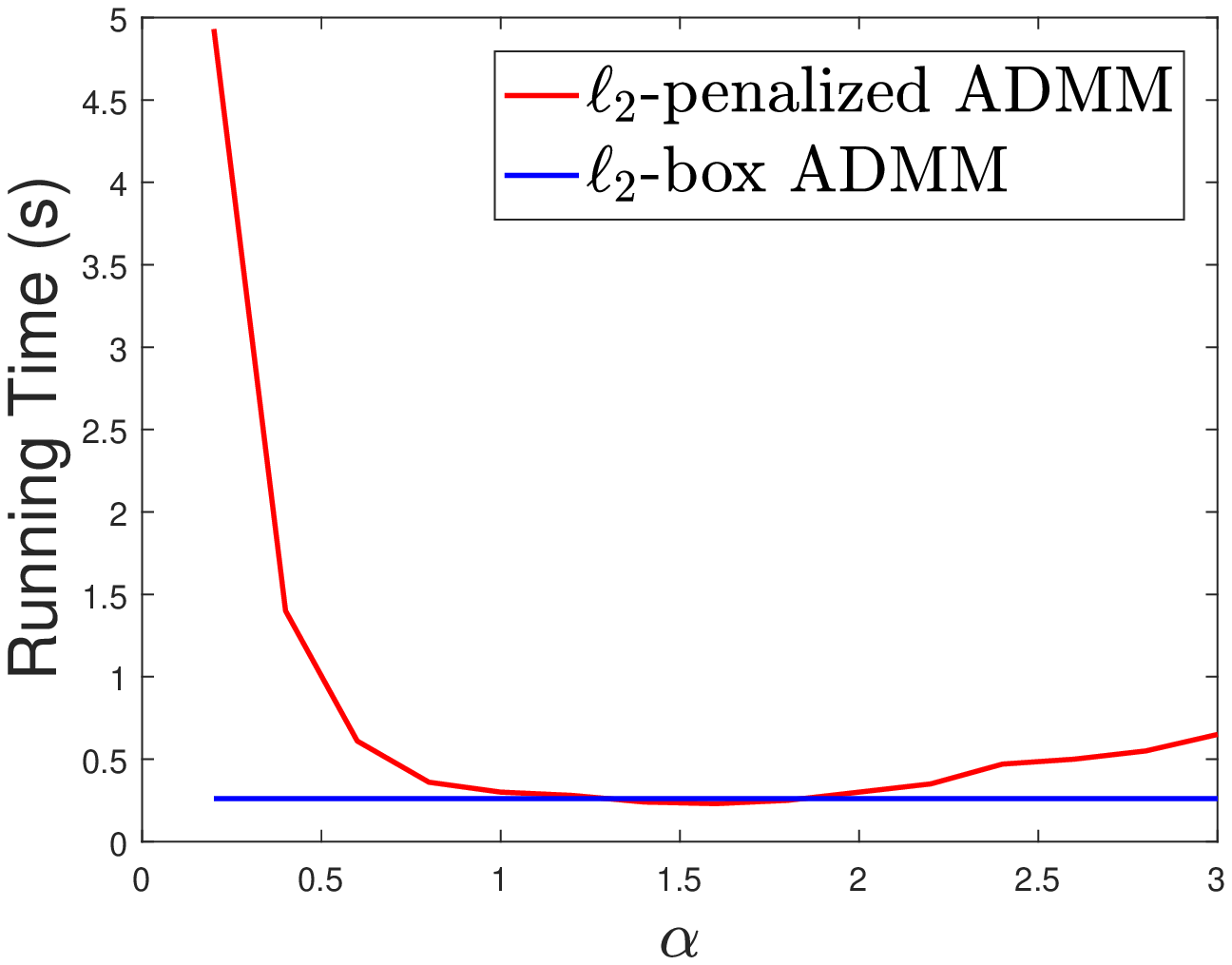}} 
\caption{WER and Running time versus $\alpha$} \label{fig:alpha} 
\end{figure}

We consider the $\ell_2$-penalized ADMM LP decoder \cite{SPBP} in our  comparison, which is reported to be the most effective and efficient among three penalty terms in \cite{SPBP}.   In this experiment, we set $\snr=1.6$dB, and the  maximum number of iterations to be 1000. The decoder is terminated    whenever 200 errors have been encountered.   We record the Word error rate (WER) and the running time of the $\ell_2$-penalized ADMM for penalty parameter value varying from 
0.24 to 5.0, and depict the results in Fig.~\ref{fig:alpha}.

%

From Fig.~\ref{fig:para:alpha_WER} and Fig.~\ref{fig:para:alpha_time}, we can see that the penalty parameter for the $\ell_2$-penalized ADMM LP decoder plays an important role in the decoding accuracy and efficiency. The algorithm becomes extremely slowly for too small or too large penalty parameter, and the WER could also be extremely large for improper parameter values.  Thus users need to select an appropriate value to achieve acceptable  performance.  As for our proposed parameter-free decoding model,  no parameter tuning task is needed for the users. 
It should be noticed that the WER and running  time of algorithm is competitive with the best performance of the $\ell_2$-penalized ADMM LP decoder. Therefor, the performance of our decoder is stable in different environment. In real situations, generally SNR cannot be estimated accurately, and it is not realistic to adjust penalty parameter in real time for broadcast transmission. These disadvantages may limit the practicability of the $\ell_2$-penalized ADMM LP decoder, while our proposed decoder does not suffer from such issue and thus is more practical in real applications.

\subsection{Performance}

\begin{figure}[htbp]
\centering 
\subfigure[Margulis]{ \label{SNR-WER} 
\includegraphics[width=1.65in]{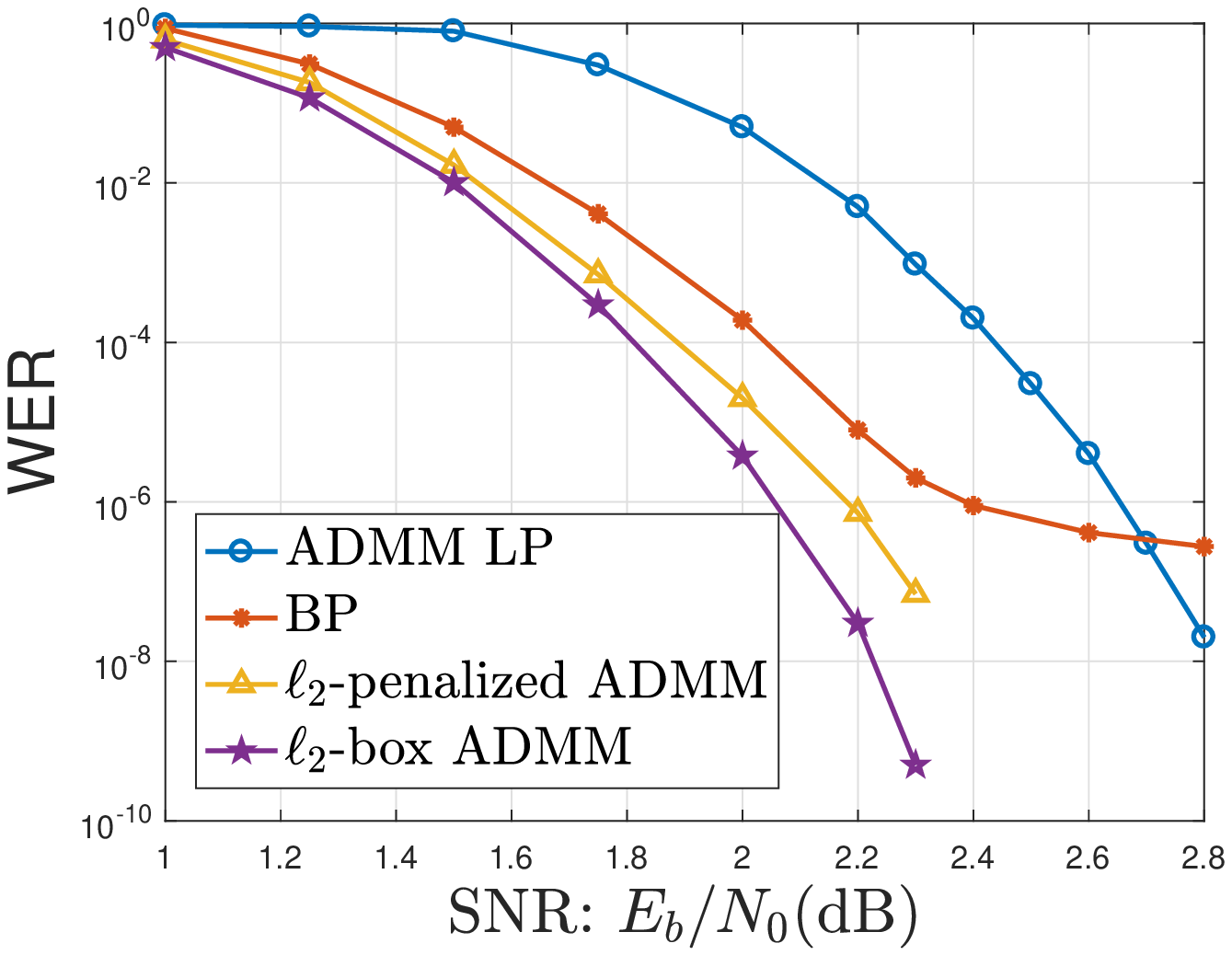}} 
\hspace{0in} 
\subfigure[WiMAX]{ \label{wimax_WER} 
\includegraphics[width=1.65in]{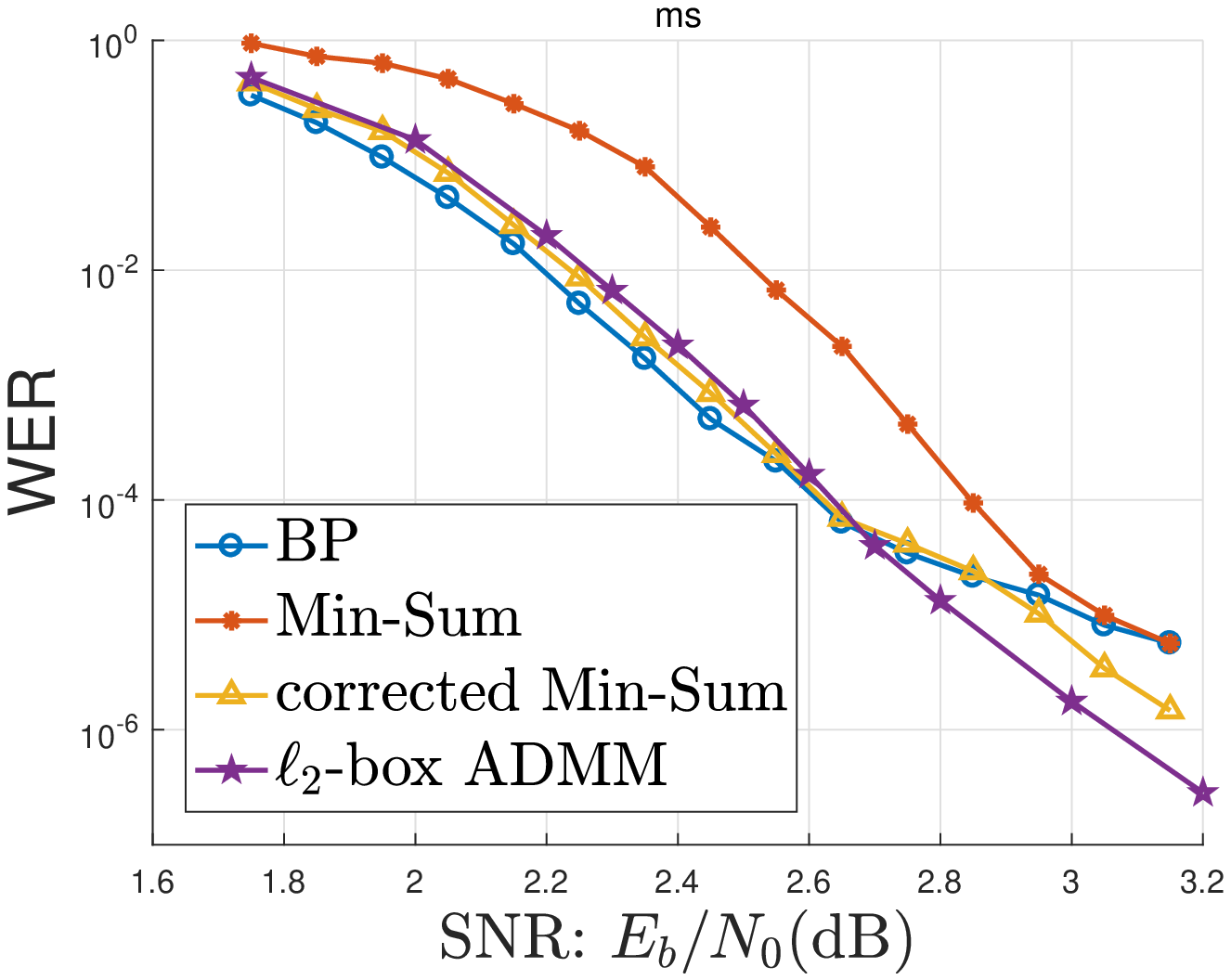}} 
\caption{WER versus different SNRs} \label{WER} 
\end{figure}

We test our proposed decoder against existing standard algorithms including 
the penalized ADMM decoder in  \cite{SPBP} and the BP decoder in \cite{DMFL}. 
The decoding model parameters for each decoder are tuned to achieve the best performance,
and the algorithms are terminated once 200 errors are encountered. The WER of each decoder is depicted  in 
Fig.~\ref{SNR-WER}.   
To show the practicality on other type of codes, we also add a comparison on $[2304, 768]$ WiMAX~\cite{wimax} in Fig.~\ref{wimax_WER}. The MS decoder and corrected MS decoder can be referred to~\cite{ryan}. The max iterations for BP decoder, MS decoder and corrected MS decoder are set as 60.

We can see that from Fig.~\ref{SNR-WER},  the overall performance of our proposed $\ell_2$-box ADMM decoder achieves lower WER  than all other contemporary decoders.  
As  the SNR turns large, the difference between  decoders become increasingly  significant, and this is especially the case when the SNR is between 1.8 and 2.4, where the error floor for BP starts from 2.  From Fig.~\ref{wimax_WER}, we can also see a significant difference when SNR is between 4 and 4.4, where the error floor for MS occurs. 
Generally the error-floor effect can be caused by weak distance spectrum of the code or the decoder’s poor ability of escaping from local minimizers. Our experiments show that the proposed decoder is more capable of avoiding early convergence to local solutions, thus can achieve better error performance. 


 \begin{table}[htbp]
   \renewcommand{\arraystretch}{1.3}
   \caption{Comparison of complexity of each decoder}
   \label{complexity}
   \centering
   \begin{tabular}{|c|c|c|c|c|}
     \hline
     \multicolumn{2}{|c|}{Decoders} & BP  & $\ell_2$-penalized & $\ell_2$-box\\
     \hline
	 \multirow{2}*{Time (s)} &SNR=2.0dB &  0.027  & 0.010   &  0.018 \\
     \cline{2-5}
     &SNR=2.5dB &  0.019  &  0.007  &  0.012 \\
     \hline
   \end{tabular}
 \end{table}

We also recorded CPU time needed for BP, $\ell_2$-penalized ADMM and $\ell_2$-box ADMM in Table~\ref{complexity}. The simulation results confirm that LP decoding outperforms BP decoding as reported in~\cite{MPAA}. It can also be observed that the greatly improved performance on WER by our proposed decoder only cost slightly additional amount of time (less than 0.008 second).

\subsection{Choice of algorithmic parameters}
\begin{figure}
\centering
\includegraphics[width=2in]{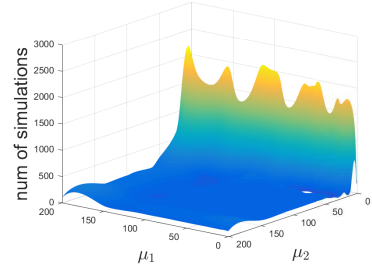}
\caption{The iterations of choice of the two parameters}
\label{2mu1-200}
\end{figure}

Our proposed decoder involves the ADMM parameters $\mu_1$ and $\mu_2$, which may affect the efficiency of the decoder.   Fig.~\ref{2mu1-200} depicts the simulations when the decoder encounter 200 errors for $(\mu_1,\mu_2) \in [0,200]\times[0,200]$ with $\snr=1.6$dB.. We can see that for $(\mu_1,\mu_2) \in [0,200]\times[10,200]$  our proposed decoder is relatively insensitive to the choice of $\mu_1$ and $\mu_2$.


\section{Conclusion}\label{sec.conclusion}

In this paper, we have proposed a parameter-free $\ell_p$-box formulation for LDPC decoding and implemented the new formulation with an  ADMM solver. 
We have also developed an efficient ADMM-based algorithm to solve this newly formulated problem in a distributive manner. It should be emphasized that the proposed decoder can be easily applied to  various situations with different SNR and channels  since the decoding model is parameter-free.
 

\bibliographystyle{IEEEtran}


\end{document}